\documentclass[10pt, prd, aps, showpacs, twocolumn, nofootinbib]{revtex4-1}

\usepackage{amsmath,amssymb}
\usepackage{graphicx}
\usepackage{subfig}
\usepackage{caption}
\newcommand{\be}{\begin{equation}}
\newcommand{\ee}{\end{equation}}
\newcommand{\bea}{\begin{eqnarray}}
\newcommand{\eea}{\end{eqnarray}}

\newcommand{\Romatre}{Dipartimento di Matematica e Fisica, Universit\`a  Roma Tre and INFN Sezione di Roma Tre,\\ Via della Vasca Navale 84, I-00146 Rome, Italy}
\newcommand{\RomatreINFN}{Istituto Nazionale di Fisica Nucleare, Sezione di Roma Tre,\\ Via della Vasca Navale 84, I-00146 Rome, Italy}
\newcommand{\soton}{Department of Physics and Astronomy, University of Southampton,\\ Southampton SO17 1BJ, UK}
\newcommand{\Romadue}{Dipartimento di Fisica, Universit\`a di Roma ``Tor Vergata" and INFN Sezione di Tor Vergata,\\ Via della Ricerca Scientifica 1, I-00133 Roma, Italy}
\newcommand{\LaSapienza}{Dipartimento di Fisica, Universit\`a di Roma La Sapienza and INFN Sezione di Roma,\\ Piazzale Aldo Moro 5, 00185 Roma, Italy}

\begin{document}

\title{First lattice calculation of the QED corrections to leptonic decay rates}

\author{D.~Giusti}\affiliation{\Romatre}
\author{V.~Lubicz}\affiliation{\Romatre}
\author{G.~Martinelli}\affiliation{\LaSapienza}
\author{C.T.~Sachrajda}\affiliation{\soton}
\author{F.~Sanfilippo}\affiliation{\RomatreINFN}
\author{S.~Simula}\affiliation{\RomatreINFN}
\author{N.~Tantalo}\affiliation{\Romadue}
\author{C.~Tarantino}\affiliation{\Romatre}


\begin{abstract} 
The leading-order electromagnetic and strong isospin-breaking corrections to the ratio of $K_{\mu 2}$ and $\pi_{\mu 2}$ decay rates are evaluated for the first time on the lattice, following a method recently proposed. The lattice results are obtained using the gauge ensembles produced by the European Twisted Mass Collaboration with $N_f = 2 + 1 + 1$ dynamical quarks. Systematics effects are evaluated and the impact of the quenched QED approximation is estimated. Our result for the correction to the tree-level $K_{\mu 2} / \pi_{\mu 2}$ decay ratio is $-1.22\,(16) \%$ to be compared to the estimate $-1.12\,(21) \%$ based on Chiral Perturbation Theory and adopted by the Particle Data Group. 
\end{abstract}

\maketitle

\section{Introduction}
\label{sec:intro}

The determination of a number of hadronic quantities relevant for flavour physics phenomenology using lattice QCD simulations has reached such an impressive level of precision~\cite{FLAG} that both electromagnetic (e.m.) and strong isospin-breaking (IB) effects cannot be neglected.

In the past few years accurate lattice results including e.m.~and IB effects have been obtained for the hadron spectrum, as in the case of the charged/neutral mass splittings of pseudoscalar (P) mesons and baryons (see, e.g., Refs.~\cite{deDivitiis:2013xla,Borsanyi:2014jba}).
In this respect the inclusion of QED effects in lattice QCD simulations has been carried out following mainly two methods: in the first one QED is added directly to the action and QCD+QED simulations are performed at few values of the electric charge (see, e.g., Refs.~\cite{Borsanyi:2014jba,Boyle:2017gzv}), while the second one, the RM123 approach of Refs.~\cite{deDivitiis:2013xla,Giusti:2017dmp}, consists in an expansion of the lattice path-integral in powers of two {\it small} parameters: the e.m.~coupling $\alpha_{em}$ and the light-quark mass difference $(m_d - m_u) / \Lambda_{QCD}$, which are both at the level of $\approx 1 \%$.
Since it suffices to work at leading order in the perturbative expansion, the attractive feature of the RM123 method is that the small values of the two expansion parameters are factorised out, so that one can get relatively large numerical signals for the {\it slopes} of the corrections with respect to the expansion parameters. 
Moreover the slopes can be determined in isosymmetric QCD.
In this letter we adopt the RM123 method.

While the calculation of e.m.~effects in the hadron spectrum does not suffer from infrared (IR) divergences, the same is not true in the case of  hadronic amplitudes, where e.m.~IR divergences are present and cancel for well defined, measurable physical quantities only after including diagrams containing both real and virtual photons~\cite{BN37}.
This is the case, for example, for the leptonic $\pi_{\ell 2}$ and $K_{\ell 2}$ and the semileptonic $K_{\ell 3}$ decays, which play a crucial role for an accurate determination of the Cabibbo-Kobayashi-Maskawa (CKM) entries $|V_{us} / V_{ud}|$ and $|V_{us}|$~\cite{CKM}. 

The presence of IR divergences requires the development of additional strategies to those used in the  computation of e.m.~effects in the hadron spectrum. 
Such a new strategy was proposed in Ref.~\cite{Carrasco:2015xwa}, where the lattice determination of the decay rate of a charged pseudoscalar meson into either a final $\ell^\pm \nu_\ell$ pair or $\ell^\pm \nu_\ell \gamma$ state was addressed.
Although it is possible in lattice simulations to compute the e.m.~corrections due to the emission of real photons, this is not strictly necessary.
Instead, the amplitude for the emission of a real photon can be computed in perturbation theory by limiting the maximum energy of the emitted photon in the meson rest-frame, $\Delta E_\gamma$, to be small enough so that the internal structure of the decaying meson is not resolved, but larger than the experimental energy resolution\,\cite{Carrasco:2015xwa}. 
The IR divergences are independent of the structure of the hadrons (i.e.~they are universal) and cancel between diagrams containing a virtual photon (computed non-perturbatively) and those with the emission of a real photon (calculated perturbatively). 
In the intermediate steps of the calculation, however, it is necessary to introduce an IR regulator. 
We use the lattice volume $L^3$ itself as the IR regulator by working in the QED$_{\rm L}$ finite-volume formulation of QED\,\cite{Hayakawa:2008an}. 

The inclusive rate $\Gamma(P_{\ell 2})$ can be expressed as \cite{Carrasco:2015xwa}
 \bea
     \label{eq:Gamma}
     \Gamma(P_{\ell 2}) & = & \Gamma_0 + \Gamma_1^{pt}(\Delta E_\gamma) \\
     & = & \mbox{lim}_{L \to \infty} \left[ \Gamma_0(L) - \Gamma_0^{pt}(L) \right] \nonumber \\
     & + & \mbox{lim}_{\mu_\gamma \to 0} \left[ \Gamma_0^{pt}(\mu_\gamma) + 
               \Gamma_1^{pt}(\Delta E_\gamma, \mu_\gamma) \right] , \nonumber
 \eea
where the subscripts $0,1$ indicate the number of photons in the final state and the superscript {\em pt} denotes the point-like approximation for the decaying meson. 
The terms $\Gamma_0(L)$ and $\Gamma_0^{pt}(L)$ are evaluated on the lattice; both have the same IR divergences which therefore cancel in the difference. 
We use $L$ as the intermediate IR regulator and $\Gamma_0 - \Gamma_0^{pt}$ is independent of the regulator as this is removed~\cite{Lubicz:2016xro}. 
Since all momentum modes contribute to it,  $\Gamma_0(L)$ depends on the structure of P and must be computed non-perturbatively. 
In the second term on the r.h.s.\,of Eq.\,(\ref{eq:Gamma}) P is a point-like meson and both $\Gamma_0^{pt}(\mu_\gamma)$ and $\Gamma_1^{pt}(\Delta E_\gamma, \mu_\gamma)$ can be calculated directly in infinite volume in perturbation theory, using a small photon mass $\mu_\gamma$ as the intermediate IR regulator. 
Each term is IR divergent, but the sum is convergent\,\cite{BN37} and independent of the IR regulator.
The explicit perturbative calculations of $\Gamma^{pt}_0+\Gamma^{pt}_1(\Delta E_\gamma)$ and $\Gamma_0^{pt}(L)$ have been performed in Refs.~\cite{Carrasco:2015xwa} and \cite{Lubicz:2016xro}, respectively.

The inclusive decay rate (\ref{eq:Gamma}) can be written as
 \be
      \Gamma(P^\pm \to \ell^\pm \nu_\ell [\gamma]) =  \Gamma_{P}^{(tree)} \cdot \left[ 1 + \delta R_{P} \right] ~ ,
      \label{eq:master}
 \ee
where $\Gamma_{P}^{(tree)}$ is the tree-level decay rate given by
 \be
     \Gamma_{P}^{(tree)} = \frac{G_F^2}{8 \pi} |V_{q_1 q_2}|^2 m_\ell^2 \left( 1 - \frac{m_\ell^2}{M_{P}^2} \right)^2 
                                             \left[ f_{P}^{(0)} \right]^2 M_{P} ~ ,
     \label{eq:Gamma0}
 \ee
where $M_{P}$ is the physical mass of the charged P-meson, including both e.m.~and strong IB corrections. The superscript {\footnotesize$(0)$} on a physical quantity denotes that it has been calculated in isosymmetric QCD (without QED). 
The  P-meson decay constant, $f_{P}^{(0)}$  is defined by: 
 \be
      A_{P}^{(0)} \equiv \langle 0 | \bar{q}_2 \gamma_0 \gamma_5 q_1 | P^{(0)} \rangle \equiv f_{P}^{(0)} M_{P}^{(0)} \, .
      \label{eq:APS0}
 \ee
In Eq.~(\ref{eq:master}) the quantity $\delta R_{P}$ encodes the leading-order e.m.~and strong IB corrections to the tree-level decay rate and its evaluation is described in Ref.~\cite{Carrasco:2015xwa}.
Its value depends on the prescription used for the separation between the QED and QCD corrections, while the quantity  $[f_{P}^{(0)}]^2 (1 + \delta R_{P})$ is prescription free.
In this work we adopt the prescription in which the renormalized couplings and quark masses in the full theory and in isosymmetric QCD coincide in the $\overline{\rm MS}$ scheme at a scale of 2\,\mbox{GeV}~\cite{deDivitiis:2013xla,Giusti:2017dmp} (see the supplemental material).

In this letter we focus on the ratio of the inclusive decay rates of kaons and pions into muons, namely
\be
      \label{eq:KPi}
      \frac{\Gamma(K_{\mu 2})}{\Gamma(\pi_{\mu 2})} = \left| \frac{V_{us}}{V_{ud}} \frac{f_K^{(0)}}{f_\pi^{(0)}} \right|^2 
         \frac{M_\pi^3}{M_K^3} \left( \frac{M_K^2 - m_\mu^2}{M_\pi^2 - m_\mu^2} \right)^2 
         \left( 1 + \delta R_{K \pi} \right) ,
 \ee
where $\delta R_{K \pi} \equiv \delta R_K - \delta R_\pi$.
Using the gauge ensembles generated by the European Twisted Mass Collaboration (ETMC) with $N_f = 2 + 1 + 1$ light, strange and charm sea quarks~\cite{Baron:2010bv,Baron:2011sf}, we have calculated $\delta R_{K \pi}$, which, together with a lattice computation of $f_K^{(0)} / f_\pi^{(0)}$, allows us to determine $|V_{us} / V_{ud}|$
from the ratio in Eq.\,(\ref{eq:KPi}). 

The quantity $\delta R_{K \pi}$ is less sensitive to various uncertainties than the individual terms $\delta R_K$ and $\delta R_\pi$.
Three main features help to reduce the systematic uncertainties in $\delta R_{K \pi}$:~(i) In $\Gamma_0(L)$ all the terms up to ${\cal{O}}(1/L)$ are ``universal'', i.e.~independent of the structure of the decaying hadron~\cite{Lubicz:2016xro}.
The residual, structure-dependent (SD) finite volume effects (FVEs) start at order ${\cal{O}}(1/L^2)$ and are found to be much milder in the case of $\delta R_{K \pi}$ (see section \ref{sec:FVE}). 
(ii) The matching of the bare lattice weak operator with the one renormalised using $W$-regularization generates a mixing of operators of different chiralities when discretisations based on Wilson fermions, which break the chiral symmetry (such as twisted mass used here) are used. The mixing has been calculated only at order ${\cal{O}}(\alpha_{em} \alpha_s^0)$~\cite{Carrasco:2015xwa}, but its effects cancel out in the difference $\delta R_{K \pi}$.~(iii) Within SU(3) Chiral Perturbation Theory (ChPT) the effects of the sea-quark electric charges depend on unknown low-energy constants (LECs) starting at next-to-leading-order (NLO) for $\delta R_K$ and $\delta R_\pi$, but only at NNLO for $\delta R_{K \pi}$~\cite{Bijnens:2006mk}. Thus, the uncertainty due to the quenched QED (qQED) approximation, adopted in this work, is expected to be smaller for $\delta R_{K \pi}$.

Since the experimental rates $\Gamma(K_{\mu 2})$ and $\Gamma(\pi_{\mu 2})$ are inclusive, SD contributions to the real photon emission should be included.
According to the ChPT predictions of Ref.~\cite{Cirigliano:2007ga}, however, these contributions are negligible in the case of both kaon and pion decays into muons, while the same does not hold as well in the case of final electrons (see Ref.~\cite{Carrasco:2015xwa}).
This important finding will be investigated by an ongoing dedicated lattice study on the real photon emission amplitudes in light and heavy P-meson leptonic decays.

After extrapolating our lattice data to the physical pion mass and to the continuum and infinite volume limits, the main result of the present work is
\be
      \label{eq:KPi_result}
      \delta R_{K \pi}^{phys} = - 0.0122 \pm 0.0016\,,
 \ee
where the uncertainty includes both statistical and systematic errors, including an estimate of the uncertainty due to the QED quenching.
 Our result (\ref{eq:KPi_result}) can be compared with the current estimate $\delta R_{K \pi}^{phys} = - 0.0112\,(21)$ from Refs.~\cite{Rosner:2015wva,Cirigliano:2011tm} adopted by the PDG~\cite{PDG}.

\section{Details of the simulation}
\label{sec:simulations}

The gauge ensembles used in this work were generated by ETMC with $N_f = 2 + 1 + 1$ dynamical quarks and used in Ref.~\cite{Carrasco:2014cwa} to determine the up, down, strange and charm quark masses.
The main parameters of the simulations are collected in the supplemental material.
We employ the Iwasaki action \cite{Iwasaki:1985we} for gluons and the Wilson Twisted Mass Action \cite{Frezzotti:2000nk, Frezzotti:2003xj, Frezzotti:2003ni} for sea quarks. 
In the valence sector we adopt a non-unitary setup \cite{Frezzotti:2004wz} in which the strange quark is regularized as an Osterwalder-Seiler fermion \cite{Osterwalder:1977pc}, while the up and down quarks have the same action as the sea.
Working at maximal twist such a setup guarantees an automatic ${\cal{O}}(a)$-improvement \cite{Frezzotti:2003ni, Frezzotti:2004wz}.
The two valence quarks in the P-meson are regularized with opposite values of the Wilson $r$-parameter in order to guarantee that discretization effects on the P-meson mass are of order $\mathcal{O}(a^2 \mu \, \Lambda_{QCD})$.
The lepton $\ell$ is a free twisted-mass fermion with mass $m_\ell = m_\mu = 105.66$\,MeV~\cite{PDG}. 
The neutrino is simply considered to be a free fermion field.

In this work we make use of the bootstrap samplings generated for the input parameters of the quark mass analysis of Ref.\,\cite{Carrasco:2014cwa}.
There, eight branches of the analysis were adopted differing in: (i) the continuum extrapolation, adopting for the matching of the lattice scale either the Sommer parameter $r_0$ or the mass of a fictitious P-meson made up of two valence strange(charm)-like quarks;  (ii) the chiral extrapolation performed with fitting functions chosen to be either a polynomial expansion or a ChPT Ansatz in the light-quark mass; and (iii) the choice between the methods M1 and M2, which differ by ${\cal{O}}(a^2)$ effects, used to determine the mass renormalization constant $Z_m = 1 / Z_P$ in the RI'-MOM scheme.

\section{Evaluation of the amplitudes}
\label{sec:master}

Following Ref.~\cite{Carrasco:2015xwa} the quantity $\delta R_K - \delta R_\pi$ is given by
 \bea
    \delta R_{K \pi} & = & 2 \frac{\delta A_K}{A_K^{(0)}} - 2 \frac{\delta M_K}{M_K^{(0)}} + 
                                       \delta \Gamma_K^{(pt)}(\Delta E_\gamma) \nonumber \\
                             &&\hspace{0.3cm}  -2 \frac{\delta A_\pi}{A_\pi^{(0)}} + 2 \frac{\delta M_\pi}{M_\pi^{(0)}} - 
                                      \delta \Gamma_\pi^{(pt)}(\Delta E_\gamma) ~ ,
    \label{eq:RPS}
 \eea
where $\delta \Gamma_{P}^{(pt)}(\Delta E_\gamma)$ represents the ${\cal{O}}(\alpha_{em})$ correction to the tree-level decay rate for a point-like meson and can be read off from Eq.~(51) of Ref.~\cite{Carrasco:2015xwa}, while $\delta A_{P}$ and $\delta M_{P}$ are the e.m.~and IB corrections to the weak amplitude and mass of the P-meson, respectively.

\begin{figure}[htb!]
\subfloat[]{\raisebox{0.220\totalheight}{\includegraphics[scale=0.220]{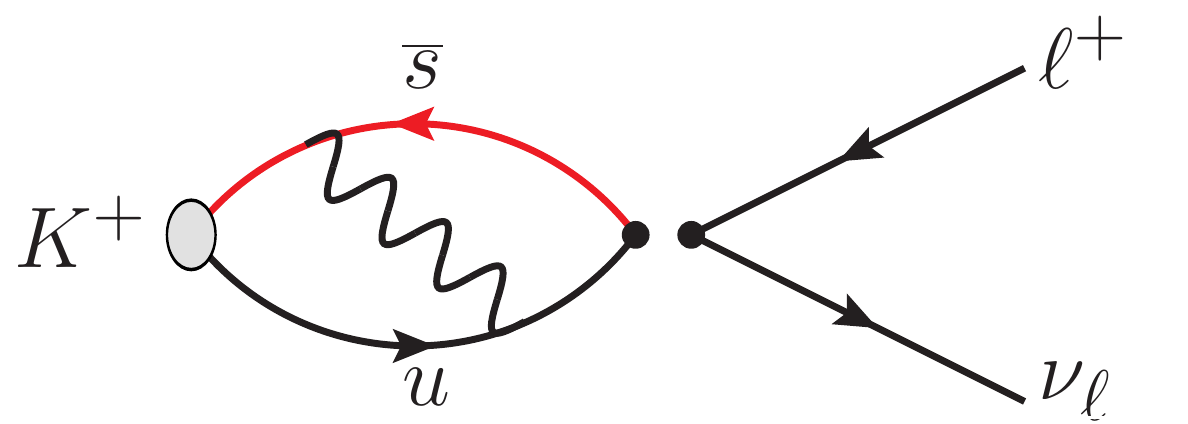}}}~
\subfloat[]{\raisebox{0.220\totalheight}{\includegraphics[scale=0.220]{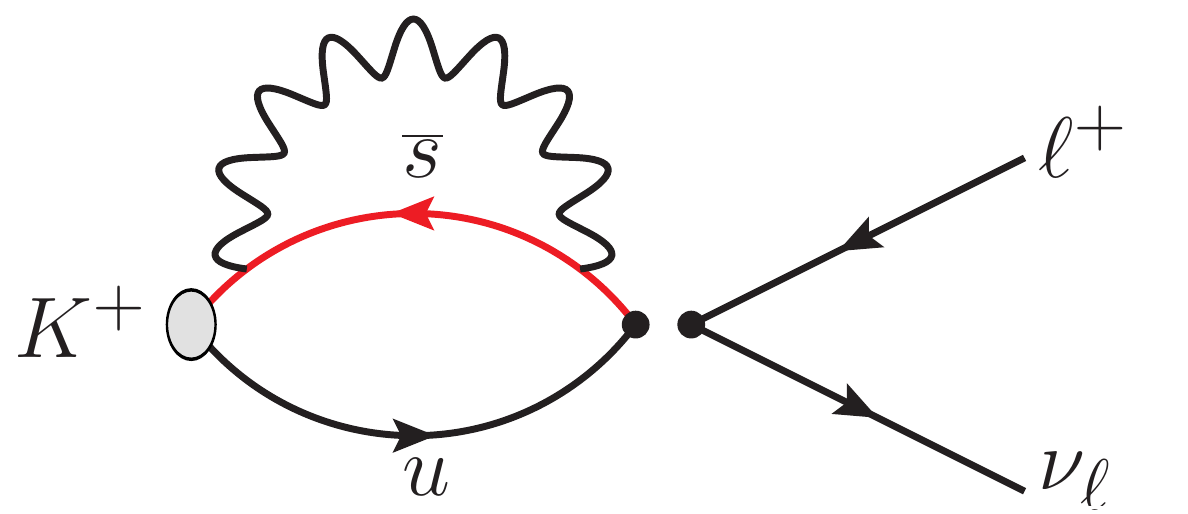}}}~
\subfloat[]{\includegraphics[scale=0.220]{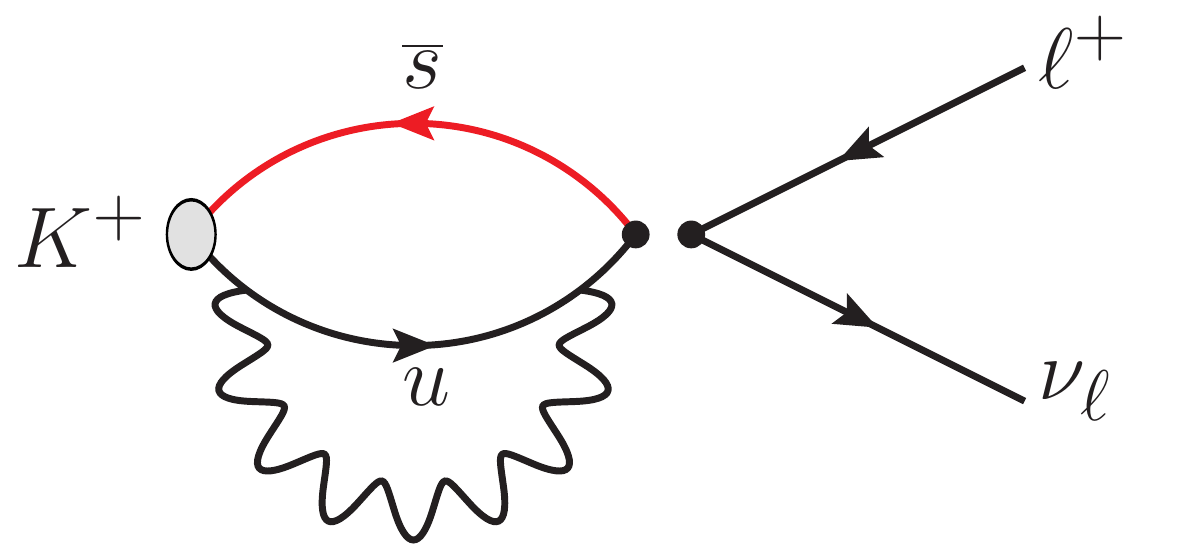}}\\[-1pt]
\subfloat[]{\raisebox{0.220\totalheight}{\includegraphics[scale=0.220]{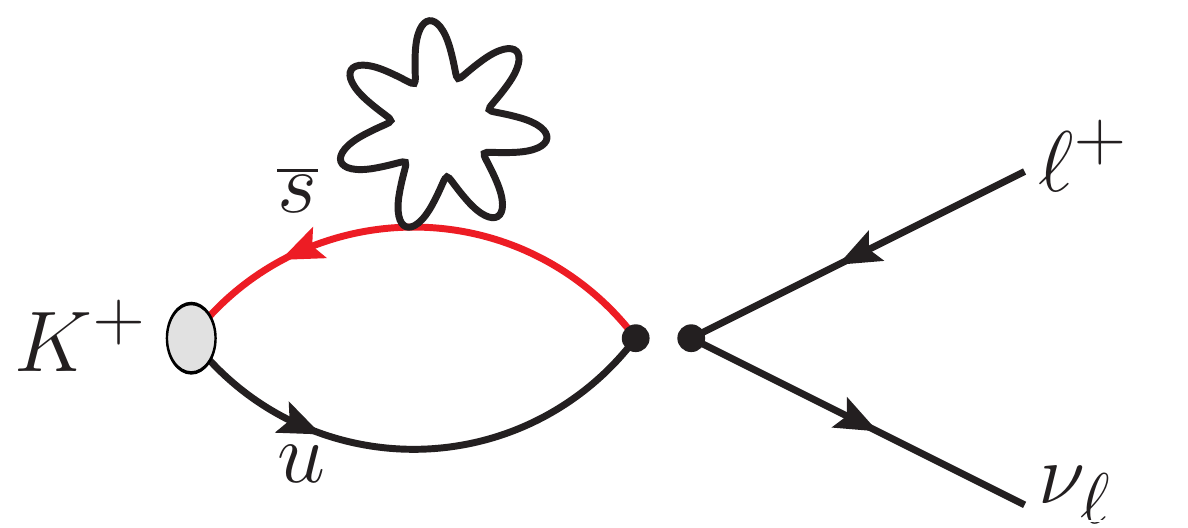}}}~
\subfloat[]{\includegraphics[scale=0.220]{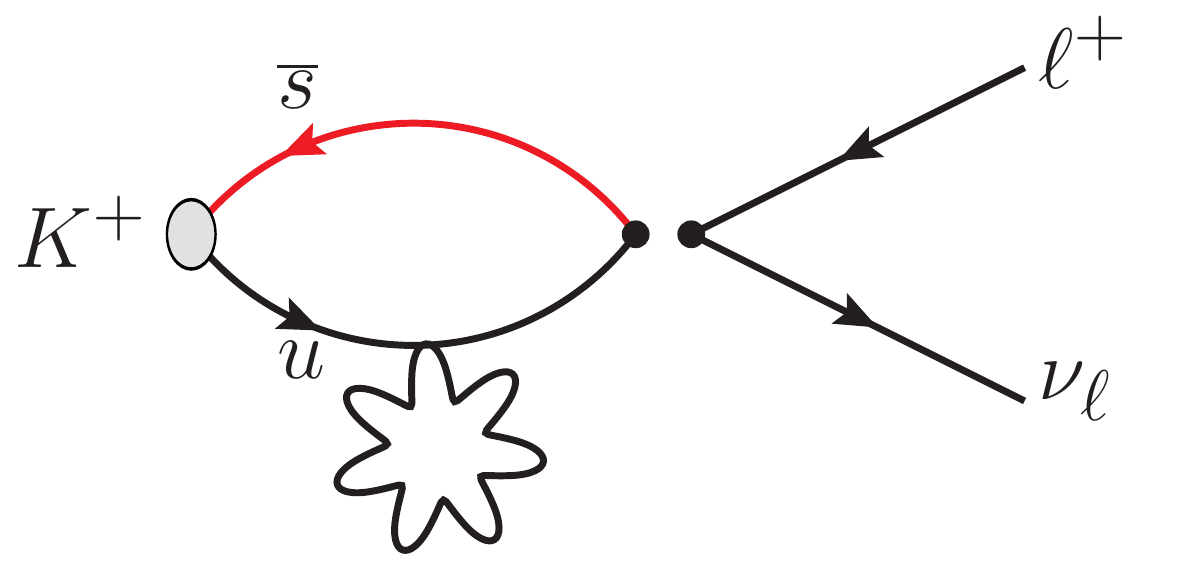}}
\vspace{-0.35cm}
\caption{\it \footnotesize Connected diagrams contributing at ${\cal{O}}(\alpha_{em})$ to the $K^+ \to \ell^+ \nu_\ell$ decay amplitude, in which the photon is attached to quark lines: (a) exchange, (b,c) self-energy and (d,e) tadpole diagrams.\hspace*{\fill}}
\label{fig:diagrams_1}
\vspace{-0.10cm}
\subfloat[]{\includegraphics[scale=0.220]{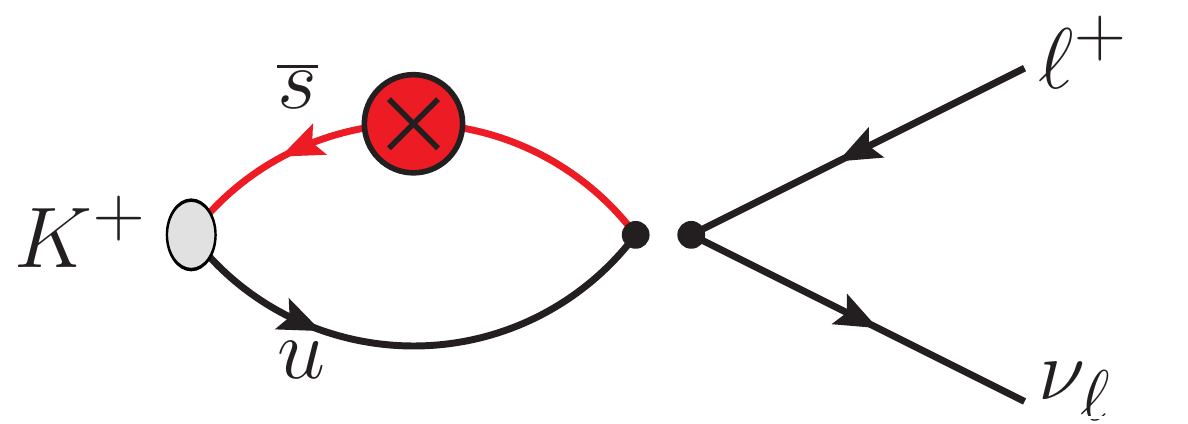}}~
\subfloat[]{\includegraphics[scale=0.220]{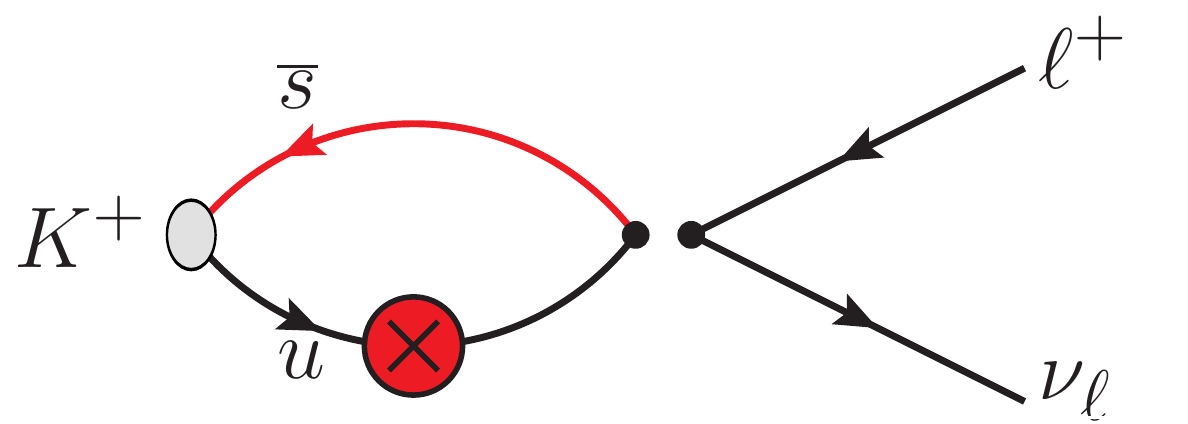}}
\vspace{-0.35cm}
\caption{\it \footnotesize Connected diagrams contributing at ${\cal{O}}(\alpha_{em})$ to the $K^+ \to \ell^+ \nu_\ell$ decay amplitude corresponding to the insertion of the pseudoscalar density related to the e.m.~shift of the critical mass, $\delta m_f^{\rm crit}$, determined in Ref.~\cite{Giusti:2017dmp}.\hspace*{\fill}}
\label{fig:diagrams_2}
\vspace{-0.10cm}
\subfloat[]{\includegraphics[scale=0.220]{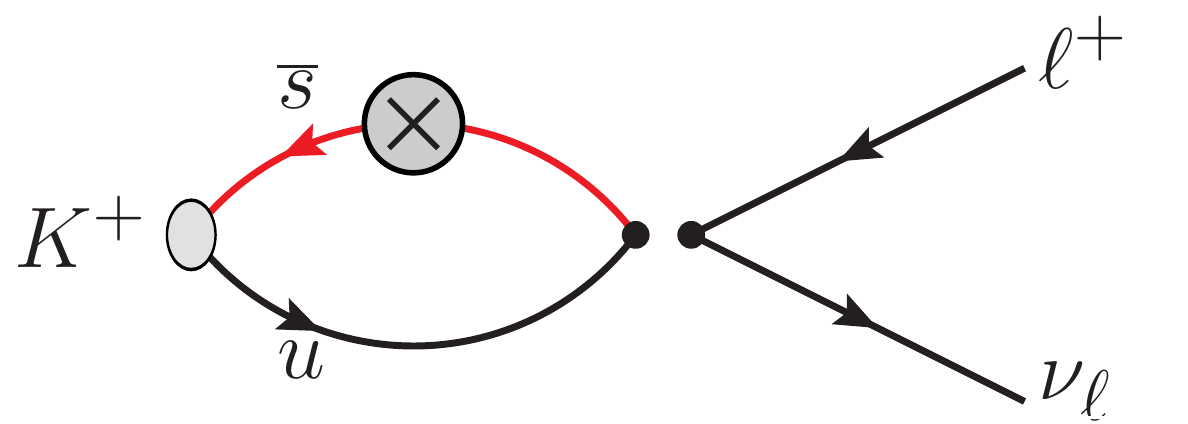}}~
\subfloat[]{\includegraphics[scale=0.220]{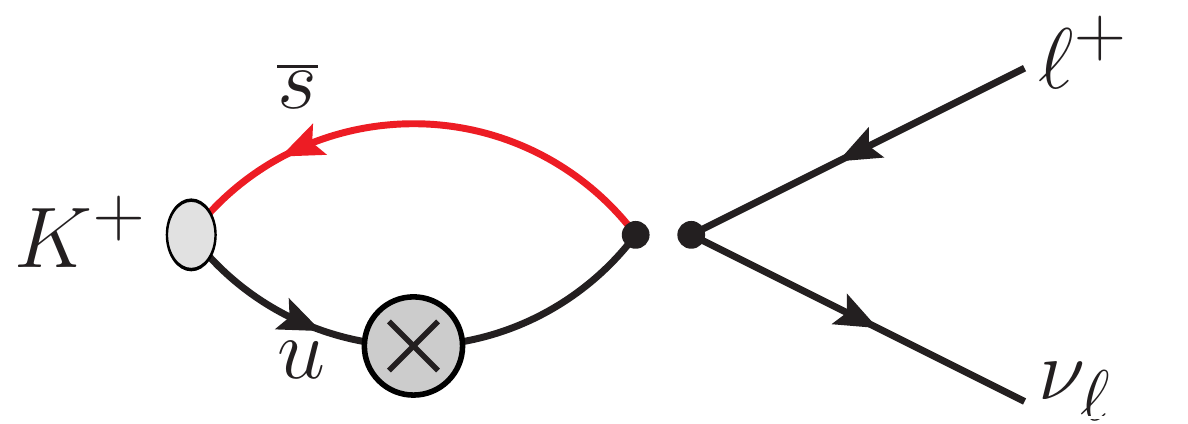}}
\vspace{-0.35cm}
\caption{\it \footnotesize Connected diagrams contributing at ${\cal{O}}(\alpha_{em})$ and ${\cal{O}}(m_d - m_u)$ to the $K^+ \to \ell^+ \nu_\ell$ decay amplitude related to the insertion of the scalar density (see Ref.~\cite{Giusti:2017dmp}).\hspace*{\fill}}
\label{fig:diagrams_3}
\vspace{-0.10cm}
\subfloat[]{\includegraphics[scale=0.220]{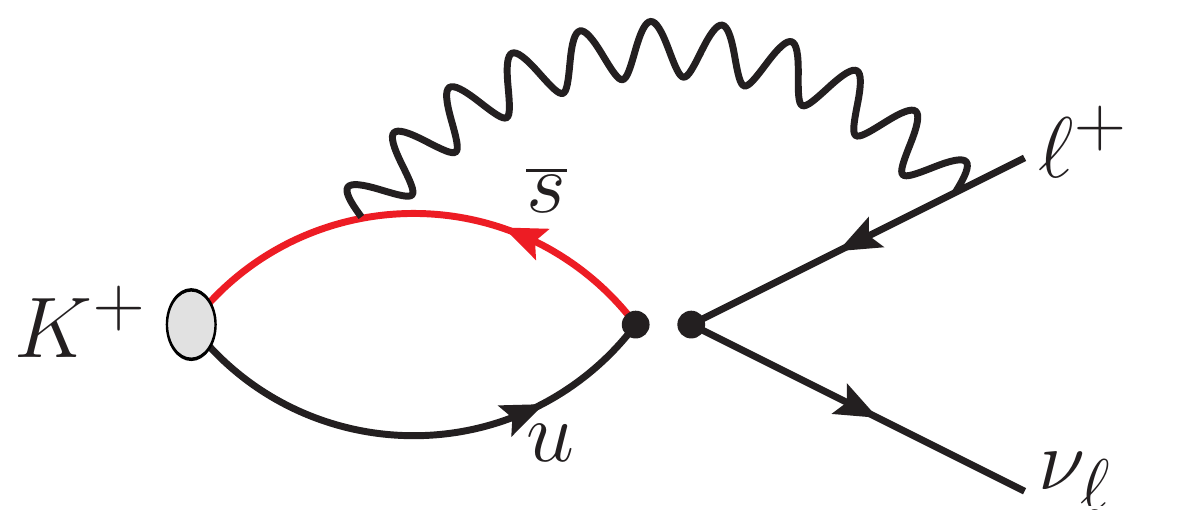}}~
\subfloat[]{\includegraphics[scale=0.220]{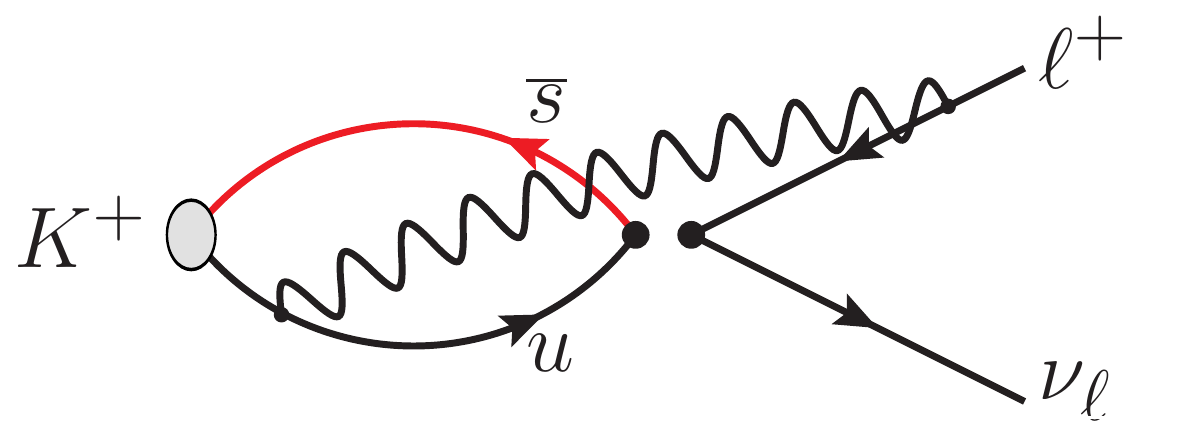}}~
\subfloat[]{\includegraphics[scale=0.220]{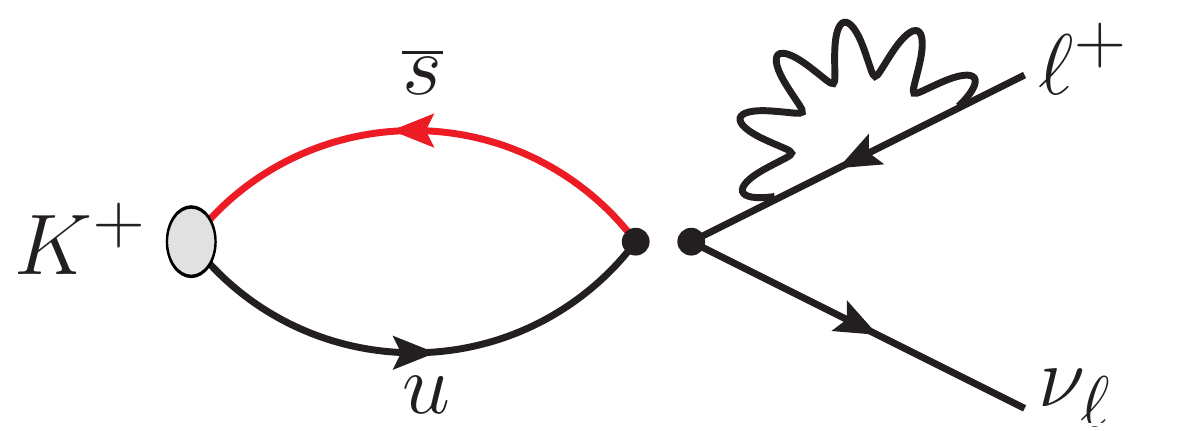}}
\vspace{-0.35cm}
\caption{\it \footnotesize Connected diagrams contributing at ${\cal{O}}(\alpha_{em})$ to the $K^+ \to \ell^+ \nu_\ell$ decay amplitude corresponding to photon exchanges involving the final-state lepton.\hspace*{\fill}}
\label{fig:diagrams_4}
\end{figure}

Within the qQED approximation the evaluation of $\delta A_{P}$ and $\delta M_{P}$ requires the evaluation of only the connected diagrams shown in Figs.~\ref{fig:diagrams_1}\,-\,\ref{fig:diagrams_4} for $K_{\ell 2}$ decays.
The corrections $\delta A_{P}$ and $\delta M_{P}$ can be written as
 \bea
      \label{eq:deltaAPS}
      \delta A_{P} & = & \delta A_{P}^{QCD} + \sum_{i = J, T, P , S} \delta A_{P}^i + \delta A_{P}^\ell~  , \\
      \label{eq:deltaMPS}
      \delta M_{P} & = & \delta M_{P}^{QCD} + \sum_{i = J, T, P , S} \delta M_{P}^i ~  ,
 \eea
where $\delta A_{P}^{QCD}$ ($\delta M_{P}^{QCD}$) represents the strong IB corrections corresponding to the diagrams of Fig.~\ref{fig:diagrams_3}, while the other terms are QED corrections coming from the insertions of the e.m.~current and tadpole operators, of the pseudoscalar and scalar densities (see Refs.~\cite{deDivitiis:2013xla,deDivitiis:2011eh}).

In Eqs.~(\ref{eq:deltaAPS}-\ref{eq:deltaMPS}) the term $\delta A_{P}^J$ ($\delta M_{P}^J$) is generated by the diagrams of Fig.~\ref{fig:diagrams_1}(a-c), $\delta A_{P}^T$ ($\delta M_{P}^T$) by the diagrams of Fig.~\ref{fig:diagrams_1}(d-e), $\delta A_{P}^P$ ($\delta M_{P}^P$) by the diagrams of Fig.~\ref{fig:diagrams_2}(a-b) and $\delta A_{P}^S$ ($\delta M_{P}^S$) by the diagrams of Fig.~\ref{fig:diagrams_3}(a-b).
The term $\delta A_{P}^{\ell}$ corresponds to the photon exchange between the quarks and the final lepton.
It arises from diagrams \ref{fig:diagrams_4}(a-b), while the diagram~\ref{fig:diagrams_4}(c) (lepton wave function renormalization) can be safely omitted, since it cancels out exactly in the difference $\Gamma_0(L) - \Gamma_0^{pt}(L)$.

The evaluation of $\delta M_{P}^{QCD}$ and the $\delta M_{P}^i$ is described in Ref.~\cite{Giusti:2017dmp}, where the quark mass difference $(m_d - m_u)(\overline{\rm MS}, 2\,\mbox{GeV}) = 2.38\,(18)$\,MeV was determined using the experimental charged and neutral kaon masses.
The terms $\delta A_{P}^{QCD}$, $\delta A_{P}^i$ and $\delta A_{P}^\ell$ are extracted from the correlators described in Ref.~\cite{Carrasco:2015xwa}.
Their numerical determination is illustrated briefly in Refs.~\cite{Lubicz:2016mpj,Simula:2017brd} and in detail in Ref.~\cite{future}.
The quality of the extraction of $\delta A_P^{\ell = \mu} / \delta A_P^{(0)}$ is illustrated in the supplemental material.

\section{Finite volume effects at \boldmath{${\cal{O}}(\alpha_{em})$}}
\label{sec:FVE}

The subtraction $\Gamma_0(L) - \Gamma_0^{pt}(L)$ makes the rate IR finite and cancels the structure-independent FVEs. 
The point-like decay rate $ \Gamma_0^{pt}(L)$ is given by  
 \be
     \Gamma_0^{pt}(L) = 2 \frac{\alpha_{em}}{4\pi} ~ Y_{P}(L) ~ \Gamma_{P}^{tree} ~ ,  
 \ee
where the factor $Y_{P}(L)$ is explicitly given by Eq.~(98) of Ref.~\cite{Lubicz:2016xro}.
Eq.~(\ref{eq:deltaAPS}) is therefore replaced by
 \be
     \delta A_{P} = \delta A_{P}^{QCD} + \sum_i \delta A_{P}^i + \delta A_{P}^\ell - 
                            \frac{\alpha_{em}}{4\pi} Y_{P}(L) \, A^{(0)}_{P} ~ ,
     \label{eq:deltaAPS_sub}
 \ee
where $Y_{P}(L)$ has the form
 \bea
    Y_{P}(L) & = & b_{IR} ~ \mbox{log}(M_{P} L) + b_0 + \frac{b_1}{M_{P} L} \nonumber \\
                  & + & \frac{b_2}{(M_{P} L)^2} + \frac{b_3}{(M_{P} L)^3} + {\cal{O}}(e^{-M_{P} L}) ~ 
       \label{eq:APS_pt}
 \eea
with the coefficients $b_j$ ($j = IR, 0, 1, 2, 3$) depending on the  dimensionless ratio $m_\ell / M_{P}$~\cite{Lubicz:2016xro}.
The important point is that the SD FVEs start only at order ${\cal{O}}(1/L^2)$, i.e.~all the terms up to ${\cal{O}}(1/L)$ in Eq.~(\ref{eq:APS_pt}) are ``universal''~\cite{Lubicz:2016xro}.
Being independent of the structure they can be computed for a point-like charged meson. 

\begin{figure}[t!]
\centering
\includegraphics[scale=0.42]{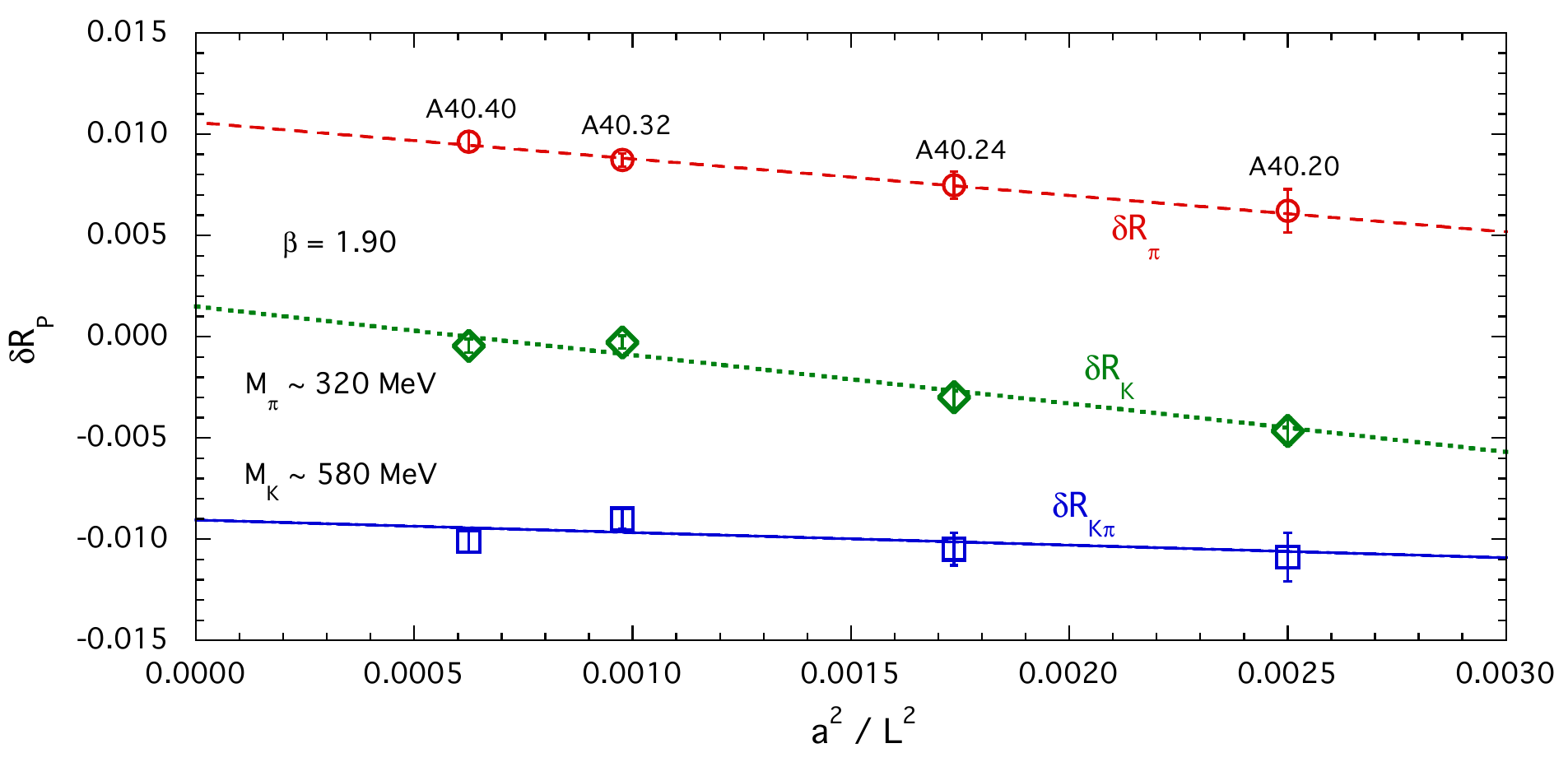}
\vspace{-0.4cm}
\caption{\it \footnotesize Results for the corrections $\delta R_\pi$, $\delta R_K$ and $\delta R_{K \pi}$ for the gauge ensembles A40.20, A40.24, A40.32 and A40.40 sharing the same lattice spacing, pion and kaon masses, but different lattice sizes (see the supplemental material). The universal FVEs, i.e.~the terms up to order ${\cal{O}}(1/L)$ in Eq.~(\protect\ref{eq:APS_pt}), are subtracted for each quantity. The lines are linear fits in $1 /L^2$. The maximum photon energy $\Delta E_\gamma$ corresponds to the inclusive case $\Delta E_\gamma = \Delta E_\gamma^{max, P} = M_{P} (1 - m_\mu^2 / M_{P}^2) / 2$.\hspace*{\fill}}
\label{fig:FVE}
\end{figure}

The FVE subtraction (\ref{eq:deltaAPS_sub}) up to order ${\cal{O}}(1/L)$ is illustrated in Fig.~\ref{fig:FVE} for $\delta R_K$, $\delta R_\pi$ and $\delta R_{K \pi}$ in the inclusive case $\Delta E_\gamma = \Delta E_\gamma^{max, P} = M_{P} (1 - m_\mu^2 / M_{P}^2) / 2$, which corresponds to $\Delta E_\gamma^{max, K} \simeq 235$\,MeV and $\Delta E_\gamma^{max, \pi} \simeq 29$\,MeV, respectively. 
It can be seen that after subtraction of the universal terms the residual FVEs are almost linear in $1 / L^2$ and $\approx 3$ times smaller in the case of $\delta R_{K \pi}$.

\section{Results for the ratio $\Gamma(K_{\ell 2}) / \Gamma(\pi_{\ell2})$}
\label{sec:results}

\begin{figure}[t!]
\centering
\includegraphics[scale=0.425]{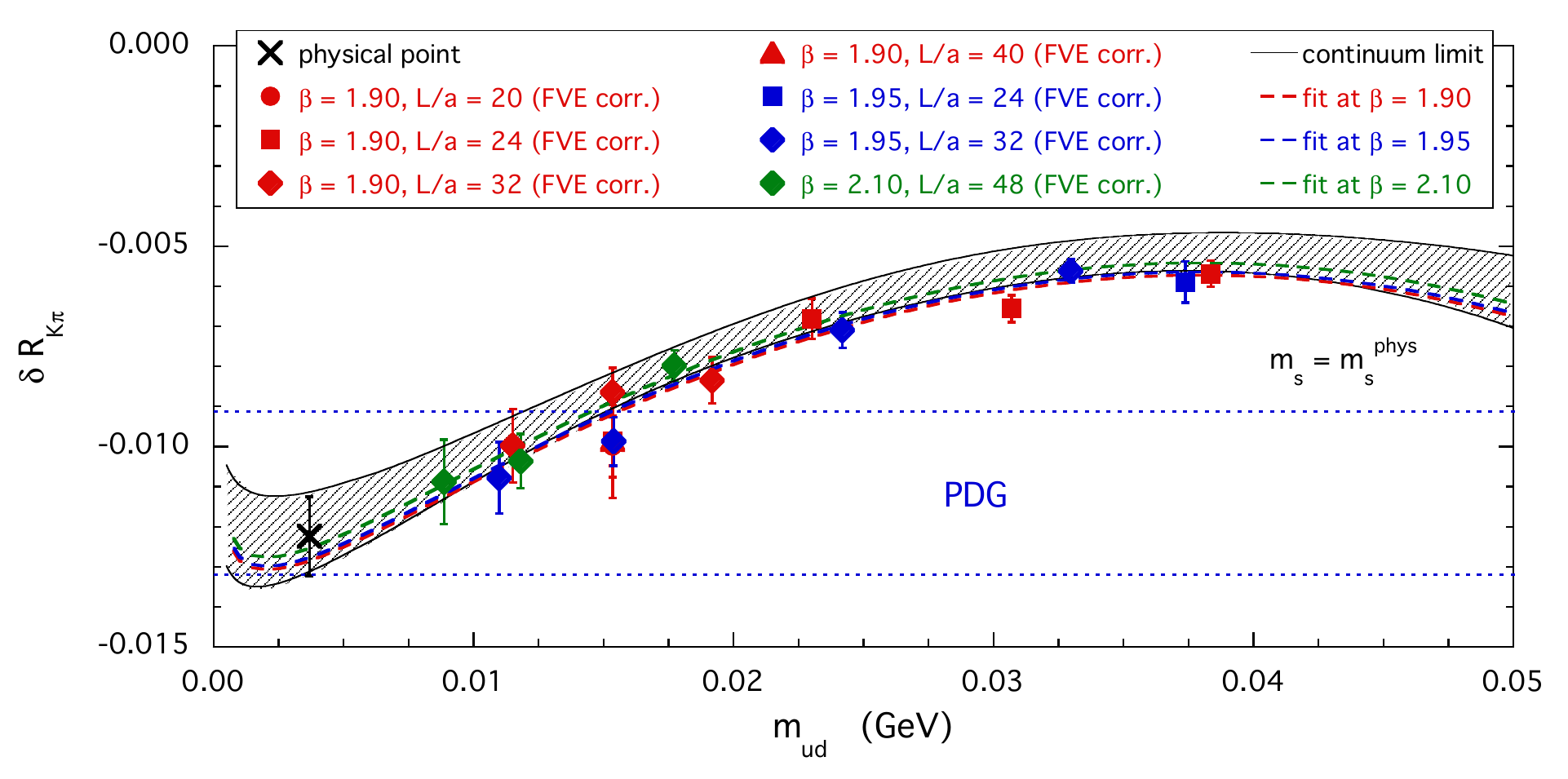}
\vspace{-0.45cm}
\caption{\it \footnotesize Results for the correction $\delta R_{K \pi}$ (Eqs.~(\ref{eq:RPS}) and (\ref{eq:deltaAPS_sub})) after the subtraction of both the universal FVEs in Eq.~(\protect\ref{eq:APS_pt}) and the residual FVEs obtained from the fitting function (\protect\ref{eq:RKPi_fit}). The dashed lines are the (central) results at each $\beta$, while the shaded area identifies the continuum limit at 1-sigma level. The cross is the extrapolated value at  $m_{ud}^{phys}(\overline{\rm MS}, 2\,\mbox{GeV}) = 3.70 (17)$\,MeV~\cite{Carrasco:2014cwa}. The blue dotted lines correspond to the value $-0.0112(21)$ from Refs.~\protect\cite{Rosner:2015wva,Cirigliano:2011tm} adopted by the PDG~\protect\cite{PDG}. Errors are statistical only.\hspace*{\fill}}
\label{fig:RKPi}
\end{figure}

The (inclusive) data for $\delta R_{K \pi}$, obtained using Eqs.~(\ref{eq:RPS}) and (\ref{eq:deltaAPS_sub}-\ref{eq:APS_pt}), are shown in Fig.~\ref{fig:RKPi}.
The ``universal'' FVEs are subtracted from the data and the combined chiral, continuum and infinite volume extrapolations are performed using the following Ansatz:
 \bea
     \delta R_{K\pi} & = & R_0 + R_\chi \mbox{log}(m_{ud}) + R_1 m_{ud}+ R_2 m_{ud}^2 + D a^2 \nonumber \\
                             & + & \frac{K_2}{L^2} \left[ \frac{1}{M_K^2} - \frac{1}{M_\pi^2} \right] + \frac{K_2^\ell}{L^2} 
                                       \left[ \frac{1}{(E_\ell^K)^2} - \frac{1}{(E_\ell^\pi)^2} \right] \nonumber \\[2mm]
                             & + & \delta \Gamma^{pt}(\Delta E_\gamma^{max, K}) - \delta \Gamma^{pt}(\Delta E_\gamma^{max, \pi}) ~ , 
     \label{eq:RKPi_fit}
 \eea
where $m_{ud}$ is the renormalized $u/d$ quark mass, $E_\ell^{P} = M_{P} (1 + m_\ell^2 / M_{P}^2) / 2$ is the lepton energy in the P-meson rest frame, and $R_{0, 1, 2}$, $D$, $K_2$ and $K_2^\ell$ are free parameters.
In Eq.~(\ref{eq:RKPi_fit}) the chiral coefficient $R_\chi$ is known~\cite{Bijnens:2006mk} and given by $R_\chi = \alpha_{em} (2 Z / 9 - 3) / 4 \pi$ in qQED, where $Z$ is obtained from the chiral limit of the ${\cal{O}}(\alpha_{em})$ correction to $M_{\pi^\pm}^2$ (i.e.~$\delta M_{\pi^\pm}^2 = 4 \pi \alpha_{em} Z f_0^2 + {\cal{O}}(m_{ud})$).
In Ref.~\cite{Giusti:2017dmp} we found $Z = 0.658\,(40)$.

Using Eq.~(\ref{eq:RKPi_fit}) we have fitted the data for $\delta R_{K\pi}$ using a $\chi^2$-minimization procedure with an uncorrelated $\chi^2$, obtaining values of $\chi^2 / \mbox{d.o.f.}$ always around $1.2$.
The uncertainties on the fitting parameters do not depend on the $\chi^2$-value, because they are obtained using the bootstrap samplings of Ref.~\cite{Carrasco:2014cwa} (see section~\ref{sec:simulations}).
This guarantees that all the correlations among the data points and among the fitting parameters are properly taken into account.
The quality of our fits is illustrated in Fig.~\ref{fig:RKPi}.

\vspace{-0.25cm} At the physical pion mass in the continuum and infinite-volume limits we obtain
 \bea
      \label{eq:RKpi_phys}
      \delta R_{K \pi}^{phys} & = & - 0.0122 ~ (10)_{stat} ~ (2)_{input} ~ (8)_{chir} ~ (5)_{FVE} \nonumber \\
                                           &     & \qquad \qquad ~ (4)_{disc} ~ (6)_{qQED} \nonumber \\ 
                                           & = & - 0.0122 ~ (16) ~ ,
 \eea
where $()_{stat}$ indicates the uncertainty induced by both the statistical errors and the fitting procedure itself;~$()_{input}$ is the error coming from the uncertainties of the input parameters of the quark-mass analysis of Ref.~\cite{Carrasco:2014cwa};  $()_{chir}$ is the difference between including or excluding the chiral logarithm in the fits (i.e. taking $R_\chi \neq 0$ or $R_\chi = 0$); $()_{FVE}$ is the difference between including ($K_2 \neq 0$ and $K_2^\ell \neq 0$) or excluding ($K_2 = K_2^\ell = 0$) the residual FVE correction in $\delta R_{K \pi}$\,\cite{future}; $()_{disc}$ is the uncertainty coming from including ($D \neq 0$) or excluding ($D = 0$) the discretisation term proportional to $a^2$; $()_{qQED}$ is our estimate of the uncertainty of the QED quenching. This is obtained using the Ansatz (\ref{eq:RKPi_fit}) with the coefficient $R_\chi$ of the chiral log fixed at the value $R_\chi = \alpha_{em} (Z - 3) / 4 \pi$, which includes the effects of the up, down and strange sea-quark charges~\cite{Bijnens:2006mk}. The change in $\delta R_{K \pi}^{phys}$ is $\simeq 0.003$, which has been already added in the central value of Eq.~(\ref{eq:RKpi_phys}). To be conservative, we use twice that value for our estimate of the qQED uncertainty.

Our result (\ref{eq:RKpi_phys}) can be compared to the value $\delta R_{K \pi}^{phys} = -0.0112\,(21)$ from Refs.~\cite{Rosner:2015wva,Cirigliano:2011tm} adopted by the PDG~\cite{PDG}.
Using in Eq.~(\ref{eq:KPi}) the experimental $K_{\mu 2}$ and $\pi_{\mu 2}$ decay rates~\cite{PDG}, we obtain 
 \be
     \left| \frac{V_{us}}{V_{ud }} \right| \frac{f_K^{(0)}}{f_\pi^{(0)}} = 0.27673 ~ (29)_{exp} ~ (23)_{th} ~ ,  
     \label{eq:VusVud}
  \ee
where the first error comes from experiments and the second one is related to the uncertainty in our result (\ref{eq:RKpi_phys}). 
Adopting the $N_f = 2 + 1 + 1$ FLAG average $f_K^{(0)} / f_\pi^{(0)} = 1.1958 ~ (26)$ \cite{FLAG} (see the supplemental material), one gets 
 \be
     \left| \frac{V_{us}}{V_{ud}} \right| = 0.23142 ~ (24)_{exp} ~ (54)_{th} ~ .
 \ee
Using the value $|V_{ud}| = 0.97417~(21)$ from super-allowed nuclear beta decays~\cite{Hardy:2014qxa}, one then has 
 \be
      |V_{us}| = 0.22544 ~ (58) ~ .
 \ee 
Thus, using $|V_{ub}| = 0.00413 ~ (49)$~\cite{PDG} the first-row CKM unitarity is confirmed to be below the per mille level, viz.
 \be
     |V_{ud}|^2 + |V_{us}|^2 + |V_{ub}|^2 = 0.99985 ~ (49) ~ .
 \ee

\section*{Acknowledgments}
We gratefully acknowledge the CPU time provided by PRACE under the project Pra10-2693 and by CINECA under the initiative INFN-LQCD123 on the BG/Q system Fermi.
V.L., G.M., S.S., C.T.~thank MIUR (Italy) for partial support under the contract PRIN 2015P5SBHT. 
G.M.~also acknowledges partial support from ERC Ideas Advanced Grant n. 267985 ``DaMeSyFla''.
C.T.S.~was partially supported by STFC (UK) grants ST/L000296/1 and ST/P000711/1.

\newpage

\onecolumngrid

\section*{Supplemental material}

Details of the calculation described in this work will be presented in Ref.~\cite{future}. 
Here, in subsection~A we collect the main parameters of the simulations performed in the isosymmetric QCD theory in Ref.~\cite{Carrasco:2014cwa}.
These simulations correspond to a prescription for the separation between QED and QCD corrections, which is different from that of Ref.~\cite{deDivitiis:2013xla} adopted in this work for the calculation of $\delta R_{K \pi}$.
We show that the two prescriptions differ only by effects which are well within the uncertainties of the input parameters of Ref.~\cite{Carrasco:2014cwa}.

In subsection~B we sketch some of the key points of the numerical analysis and illustrate the quality of the results by showing the time-dependence of the most complicated diagrams, i.e.~those in Fig.~\ref{fig:diagrams_4}(a) and (b) in which a photon is exchanged between the quarks and the final-state charged lepton.

\subsection*{A. Simulation parameters}

The main parameters of the simulations performed within isosymmetric QCD in Ref.~\cite{Carrasco:2014cwa} are collected in Table~\ref{tab:simudetails}. 

\begin{table}[htb!]
\normalsize{
\renewcommand{\arraystretch}{1.25}
\begin{center}
\begin{tabular}{||c|c|c|c|c|c|c|c|c|c||}
\hline
ensemble & $\beta$ & $V / a^4$ &$a\mu_{ud}$&$a\mu_\sigma$&$a\mu_\delta$&$N_{cf}$& $a\mu_s$ & $M_\pi$ (MeV) & $M_K$ (MeV)\\
\hline \hline
$A40.40$ & $1.90$ & $40^{3}\cdot 80$ &$0.0040$ &$0.15$ &$0.19$ &$100$& $0.02363$ & 317(12) & 576(22) \\
\cline{1-1} \cline{3-4} \cline{7-7} \cline{9-10}
$A30.32$ & & $32^{3}\cdot 64$ &$0.0030$ & & &$150$& & 275(10) & 568(22) \\
$A40.32$ & & & $0.0040$ & & & $100$ & & 316(12) & 578(22) \\
$A50.32$ & & & $0.0050$ & & &  $150$ & & 350(13) & 586(22) \\
\cline{1-1} \cline{3-4} \cline{7-7} \cline{9-10}
$A40.24$ & & $24^{3}\cdot 48 $ & $0.0040$ & & & $150$ & & 322(13) & 582(23) \\
$A60.24$ & & & $0.0060$ & & &  $150$ & & 386(15) & 599(23) \\
$A80.24$ & & & $0.0080$ & & &  $150$ & & 442(17) & 618(14) \\
$A100.24$ &  & & $0.0100$ & & &  $150$ & & 495(19) & 639(24) \\
\cline{1-1} \cline{3-4} \cline{7-7} \cline{9-10}
$A40.20$ & & $20^{3}\cdot 48 $ & $0.0040$ & & & $150$ & & 330(13) & 586(23) \\
\hline \hline
$B25.32$ & $1.95$ & $32^{3}\cdot 64$ &$0.0025$&$0.135$ &$0.170$& $150$& $0.02094$ & 259~(9) & 546(19) \\
$B35.32$ & & & $0.0035$ & & & $150$ & & 302(10) & 555(19) \\
$B55.32$ & & & $0.0055$ & & & $150$ & & 375(13) & 578(20) \\
$B75.32$ &  & & $0.0075$ & & & $~80$ & & 436(15) & 599(21) \\
\cline{1-1} \cline{3-4} \cline{7-7} \cline{9-10}
$B85.24$ & & $24^{3}\cdot 48 $ & $0.0085$ & & & $150$ & & 468(16) & 613(21) \\
\hline \hline
$D15.48$ & $2.10$ & $48^{3}\cdot 96$ &$0.0015$&$0.1200$ &$0.1385 $& $100$& $0.01612$ & 223~(6) & 529(14) \\ 
$D20.48$ & & & $0.0020$  &  &  & $100$ & & 256~(7) & 535(14) \\
$D30.48$ & & & $0.0030$ & & & $100$ & & 312~(8) & 550(14) \\
 \hline   
\end{tabular}
\end{center}
\renewcommand{\arraystretch}{1.0}
}
\caption{\it Values of the valence and sea bare quark masses (in lattice units), of the pion and kaon masses for the $N_f = 2+1+1$ ETMC gauge ensembles used in Ref.~\cite{Carrasco:2014cwa} and for the gauge ensemble, A40.40 added to improve the investigation of FVEs. 
A separation of $20$ trajectories between each of the $N_{cf}$ analysed configurations. The bare twisted masses $\mu_\sigma$ and $\mu_\delta$ describe the strange and charm sea doublet according to Ref.~\cite{Frezzotti:2003xj}. The values of the strange quark bare mass $a \mu_s$, given for each $\beta$, correspond to the physical strange quark mass $m_s^{phys}(\overline{\rm MS}, 2\,\mbox{GeV}) = 99.6 (4.3)$\,MeV and to the mass renormalization constants determined in Ref.~\cite{Carrasco:2014cwa}. The central values and errors of pion and kaon masses are evaluated using the bootstrap procedure of Ref.~\cite{Carrasco:2014cwa}.\hspace*{\fill}}
\label{tab:simudetails}
\end{table}

Three values of the inverse bare lattice coupling $\beta$ and several lattice volumes have been considered. 
For the earlier investigation of FVEs ETMC had produced three dedicated ensembles, A40.20, A40.24 and A40.32, which share the same quark masses and lattice spacing and differ only in the lattice size $L$.
To improve the present investigation we have generated a further gauge ensemble, A40.40, at a larger value of $L$.

At each lattice spacing different values of the light sea quark mass have been considered. 
The light valence and sea bare quark masses are always taken to be degenerate ($a\mu_{ud}^{sea} = a\mu_{ud}^{val} = a\mu_{ud}$). 

In Ref.~\cite{Carrasco:2014cwa} the values of the physical $u/d$ and strange quark masses, $m_{ud}^{phys}(\overline{\rm MS}, 2\,\mbox{GeV}) = 3.70 (17)$\,MeV and $m_s^{phys}(\overline{\rm MS}, 2\,\mbox{GeV}) = 99.6 (4.3)$\,MeV, as well as the values of the lattice spacing, $a = 0.0885(36)$, $0.0815(30)$, $0.0619(18)$ fm at $\beta = 1.90$, $1.95$ and $2.10$, have been determined using the following inputs for the isosymmetric QCD theory: $M_\pi^{(0)} = M_{\pi^0} = 134.98$\,MeV, $M_K^{(0)} = 494.2$\,MeV and $f_\pi^{(0)} = 130.41$\,MeV.
The first two inputs correspond to the values suggested in the FLAG reviews \cite{FLAG}, while the value of $f_\pi^{(0)}$ corresponds to the use of the experimental rate $\Gamma(\pi_{\mu 2})$, the value of $|V_{ud}|$ from Ref.~\cite{Hardy:2014qxa} and the value $\delta R_\pi = 0.0176\,(21)$ obtained in ChPT~\cite{Rosner:2015wva,Cirigliano:2011tm} and currently adopted by the PDG~\cite{PDG}. 

We will refer to the choice of the above three hadronic inputs as the FLAG/PDG prescription, which differs from that of Ref.~\cite{deDivitiis:2013xla} adopted in this work.
We now show that the differences between the two prescriptions are well within the uncertainties of the input parameters of Ref.~\cite{Carrasco:2014cwa}.

In Ref.~\cite{Giusti:2017dmp} we have calculated the leading-order QED and QCD corrections to the pion and kaon masses within the prescription of Ref.~\cite{deDivitiis:2013xla}.
Consequently, also the isosymmetric pion and kaon masses corresponding to the above prescription have been evaluated, obtaining $M_\pi^{(0)} = 134.9\,(2)$\,MeV, $M_K^{(0)} = 494.4\,(1)$\,MeV.
These values differs only very slightly from the corresponding inputs used in the FLAG/PDG prescription.

In Ref.~\cite{future} we shall provide our result for $\delta R_\pi$, which differs slightly from the one obtained in ChPT~\cite{Rosner:2015wva,Cirigliano:2011tm} and adopted by the PDG~\cite{PDG}.
Since the quantity $[f_\pi^{(0)}]^2 (1 + \delta R_\pi)$ is prescription free, the resulting change of $[f_\pi^{(0)}]^2$ turns out to be less than $0.5 \%$.
Given that $[M_\pi^{(0)} / f_\pi^{(0)}]^2 \propto m_{ud}^{phys} + {\cal{O}}([m_{ud}^{phys}]^2)$, the change of the renormalized quark mass $m_{ud}^{phys}$ is less than $0.02$\,MeV and, analogously, the change in $m_s^{phys}$ is less than $0.5$\,MeV.
Such variations of the physical $u/d$ and strange quark masses produce changes in $\delta R_\pi$ and $\delta R_{K \pi}$, which are, however, well within the statistical uncertainties, as can be easily inferred from Fig.~\ref{fig:RKPi} in the case of $\delta R_{K \pi}$.

The above findings indicate that the prescription of Ref.~\cite{deDivitiis:2013xla} and the FLAG/PDG one differ only by effects which are well within the uncertainties of the input parameters of Ref.~\cite{Carrasco:2014cwa}.
This justifies the use of the FLAG average for the ratio $f_K^{(0)} / f_\pi^{(0)}$ to get Eq.~(\ref{eq:VusVud}) as well as the comparison of our result (\ref{eq:RKpi_phys}) with the ChPT prediction of Refs.~\cite{Rosner:2015wva,Cirigliano:2011tm}.

\subsection*{B. Evaluation of $\delta A_P^\mu / \delta A_P^{(0)}$}

\begin{figure}[b!]
\centering
\includegraphics[scale=0.85]{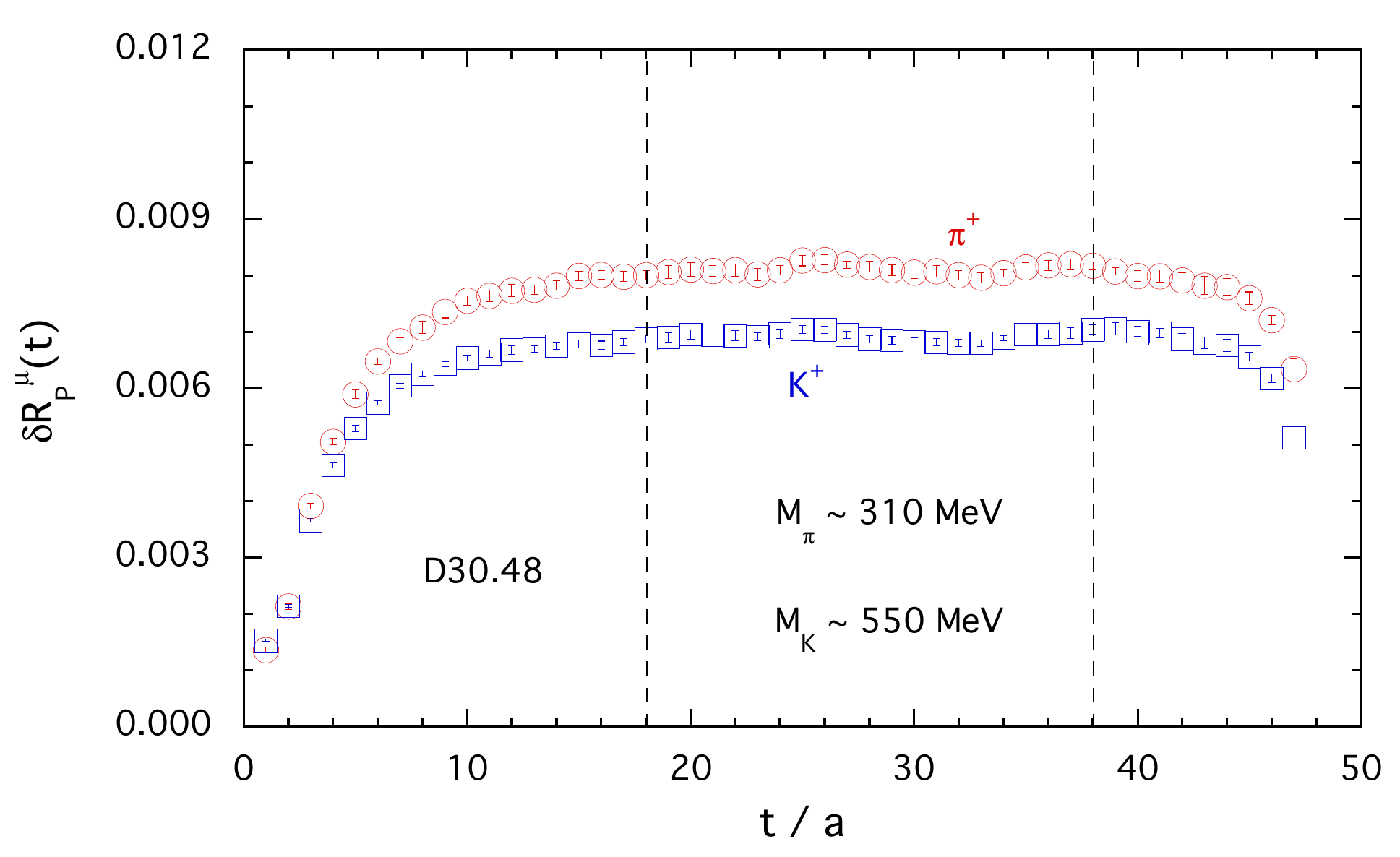}
\caption{\it Results for the ratio $\delta R_P^\mu(t)$, given by Eq.~(\ref{eq:deltaR}), in the case of kaon and pion decays to muons for the gauge ensemble D30.48. The vertical dashed lines indicate the time region used for the extraction of the amplitude ratio $\delta A_P^\mu / A_P^{(0)}$. Errors are statistical only.\hspace*{\fill}}
\label{fig:external}
\end{figure}

The evaluation of the diagrams~\ref{fig:diagrams_4}(a) and (b), corresponding to the term $\delta A_P^\ell$, starts from the correlator $\delta C_P^\ell(t)$ defined as
 \be
      \delta C_\ell(t) = \sum_{\alpha \beta} \overline{u}_{\nu_\ell \alpha}(p_{\nu_\ell})
                                 \overline{C}_1(t)_{\alpha \beta} v_{\ell \beta}(p_\ell) ~ ,
      \label{eq:external}
 \ee  
where $\overline{C}_1(t)_{\alpha \beta}$ is given by Eq.~(35) of Ref.~\cite{Carrasco:2015xwa}, while $t$ is the time distance between the P-meson source and the insertion of the weak (V-A) current.
At large time distances and for $T \to \infty$ one has
 \be
     \delta C_\ell(t) ~ _{\overrightarrow{t \gg a}} ~ \frac{Z_P^{(0)} \delta A_P^\ell}{2 M_P^{(0)}} X_P^\ell 
                                                                            e^{-M_P^{(0)} t} \,,
 \ee
where $X_P^\ell = Tr\left[ \gamma_0 (1 - \gamma_5) \ell \overline{\ell} \gamma_0 (1 - \gamma_5) \nu_\ell \overline{\nu}_\ell \right]$ is the tree-level leptonic trace evaluated on the lattice~\cite{future}.
Analogously, in the absence of the photon exchange between the quarks and the final-state lepton, the tree-level correlator $C_\ell^{(0)}(t)$ behaves at large time separations as 
 \be
     C_\ell^{(0)}(t) ~ _{\overrightarrow{t \gg a}} ~ \frac{Z_P^{(0)} A_P^{(0)}}{2 M_P^{(0)}} X_P^\ell 
                                                                           e^{-M_P^{(0)} t} ~ .
 \ee
Thus, the ratio
 \be
      \delta R_P^\ell(t) \equiv \frac{\delta C_\ell(t)}{C_\ell^{(0)}(t)} 
                                            ~ _{\overrightarrow{t \gg a}} \frac{\delta A_P^\ell}{A_P^{(0)}} 
      \label{eq:deltaR}
 \ee
should exhibit a plateau at the value $\delta A_P^\ell / A_P^{(0)}$.
The quality of the signal for $\delta R_P^{\ell = \mu}(t)$ is illustrated in Fig.~\ref{fig:external} for charged kaon and pion decays into muons in the case of the ensemble $D30.48$.

\end{document}